\documentclass[preprint,11pt]{elsarticle}

\usepackage{graphicx}
\usepackage{amssymb}

\usepackage{lineno}

\newcommand{\paraskip}{\smallskip}
\newcommand{\para}[1] {\paraskip\noindent\textbf{#1}}

\usepackage[ruled]{algorithm2e}

\usepackage{graphicx}
\usepackage{amsmath}
\usepackage[caption=false,font=footnotesize]{subfig}
\usepackage{url}
\usepackage{tabularx}
\usepackage{float}
\usepackage{enumitem}
\usepackage{listings}

\journal{Pervasive and Mobile Computing}

\begin{document}

\begin{frontmatter}


\title{Harvesting Time-Series Data from Service-Based Systems Hosted in
MANETs}



\author[label1]{Petr~Novotny}
\address[label1]{IBM T.J. Watson Research Center}
\ead{p.novotny@ibm.com}

\author[label1]{Bong~Jun~Ko}
\ead{bongjun\_ko@us.ibm.com}

\author[label2]{Alexander~L.~Wolf}
\address[label2]{University of California, Santa Cruz}
\ead{alw@acm.org}

\begin{abstract}

We are concerned with reliably harvesting data collected from
service-based systems hosted on a \emph{mobile ad hoc network}
(MANET). More specifically, we are concerned with time-bounded and
time-sensitive \emph{time-series} monitoring data describing the state
of the network and system.  The data are harvested in order to perform
an analysis, usually one that requires a global view of the data taken
from distributed sites. For example, network- and application-state
data are typically analysed in order to make operational and
maintenance decisions.
MANETs are a challenging environment in which to harvest monitoring
data, due to the inherently unstable and unpredictable connectivity
between nodes, and the overhead of transferring data in a wireless
medium. These limitations must be overcome to support time-series
analysis of perishable and time-critical data.
We present an \emph{epidemic, delay tolerant, and intelligent method}
to efficiently and effectively transfer time-series data between the
mobile nodes of MANETs. The method establishes a network-wide
synchronization overlay to transfer increments of the data over
intermediate nodes in periodic cycles. The data are then accessible
from local stores at the nodes. 
We implemented the method in Java~EE
and present evaluation on a run-time dependence discovery method for Web
Service applications hosted on MANETs, and comparison to other four methods demonstrating that our method performs significantly better in both data availability and network overhead. 

\end{abstract}

\begin{keyword}
Peer-to-peer protocols \sep
Mobile ad hoc networks \sep
MANET \sep
Availability \sep
Fault Tolerance \sep
Delay tolerant networks \sep
Time series \sep
Network-wide synchronization overlay



\end{keyword}

\end{frontmatter}


\section{Introduction}
  \label{sec:intro}

Mobile ad hoc networks (MANETs) are used to establish communication in
difficult environments such as search and rescue, forest-fire
fighting, and war zones. The increasing sophistication of end-user
devices has led to an increase in the richness and complexity of the
systems deployed on MANETs, including those structured as
interconnected and interdependent (micro)services, what are referred
to as \emph{service-based
  systems}~\cite{Esfahani+:TAAS:2016,Jalaparti+:SIGCOMM:2013,Lund+:IEEE-CM:2007,Lund+:IEEE-CM:2010}. Our
interest is in developing methods to help manage those systems in
MANET deployment environments.

The foundation of any such management method is the collection of
run-time data describing the ``health'' state of the system. In
service-based systems designed for and deployed in MANETs, these data
change frequently, as the system attempts to adapt to the dynamicity
of the underlying network. For example, the bindings between services
can dynamically change according to the status of the nodes, links,
and paths.

Time-varying, local data are observed by \emph{monitors} situated at
the nodes of the network and captured as a \emph{time series}. The
time-series data are later \emph{harvested} from the nodes and
presented to a management element located somewhere in the network so
that it can perform a global state analysis, such as fault
identification~\cite{Novotny+:IEEE-TNSM:2015,5691315}, service
discovery and composition~\cite{Issarny+:Springer-JISA:2011}, or
service
re-placement~\cite{Novotny+:IEEE-MCC:2015,Silvestri+:ICCCN:2015}. As
it turns out, harvesting is an especially difficult problem in MANETs
due to the inherent resource limitations of wireless devices and the
fact that mobility can lead to network instabilities,
asymmetries,\footnote{An asymmetric network is one in which there is a
  path from node $n$ to node $m$, but no path from $m$ to
  $n$. Wireless media are subject to such communication asymmetries,
  sometimes even between neighbors.}  and partitions resulting from
relative node movements (e.g., going in and out of range) and link
properties (e.g., directionality, interference, and
noise)~\cite{Khandani+:IEEE-TWC:2008}.

\begin{figure}[t]
\centering

\includegraphics[width=0.6\columnwidth]{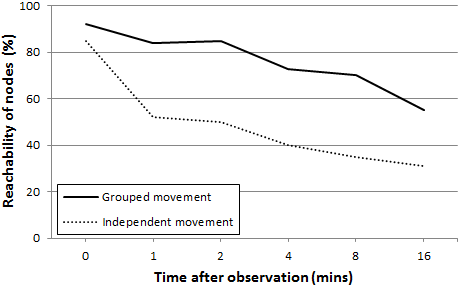}

\caption{Reachability of MANET nodes in two scenarios.}
  \label{fig:reachability}

\end{figure}

To appreciate the seriousness of the problem, consider
Figure~\ref{fig:reachability}, which is a preview of experimental
results presented in this paper. The figure shows how the reachability
of a node, and therefore access to the data stored on it, can degrade
with time in MANET environments. The experiment examines two different
mobility behaviors under the same hypothetical service-based system.
One behavior is characterized by random, independent movements of
nodes, while the other is characterized by collective, grouped
movements. The experiment models a management element, residing at
some node in the network, needing access to the monitoring data stored
on a specific subset of nodes in the network. Approximately 90\% of
those nodes are reachable at time zero. As time progresses and nodes
move about, more nodes in that subset become unreachable, meaning that
less monitoring data are available to the management
element. Interestingly, we can see that random node movements can
cause more problems than grouped node movements.

The general problem of harvesting distributed data is, of course, not
new. Many mature solutions exist in the domain of wired, stable
networks, where brute-force techniques exploit the high capacity,
reliable network environment (e.g., IBM's Tivoli). For example, the
data can be continuously streamed to one or more repositories, or the
individual nodes can be reliably contacted to provide their data on
demand.

In the dynamic, unreliable, and resource-limited domain of MANETs,
however, the key issue is guaranteeing the availability of relevant
and timely data whenever and wherever it is needed, even if a node
whose data is of interest is unreachable. Existing mechanisms for data
dissemination in MANETs either \emph{ignore the problem} of network
asymmetries and partitioning, \emph{assume prior knowledge} of the
intended destination of the data, or \emph{focus on
single aggregated values} rather than the time-series data required
for system management tasks (see Section~\ref{sec:related}).

In general, then, the harvesting method must be sensitive to the
properties of time-series data and management tasks in MANETs. The
data are \emph{time bounded}, meaning that only some portion of a
monitor's stored data will be relevant to a given task, and \emph{time
  sensitive}, meaning that their utility in a task degrades over
time. Moreover, only data from some subset of the monitors may be of
interest to a task. Finally, network instability and dynamics mean
that the location of the management node within the network topology
cannot be anticipated. In fact, the monitors may be unaware of
whether, which, and where their data might eventually be used.

Conceptually, our idea is to spread the data in an \emph{epidemic},
\emph{delay tolerant}, and \emph{intelligent} fashion, trading extra
communication and storage for increased availability in the presence
of node reachability problems. In particular, the method is:

\begin{itemize}[leftmargin=*,topsep=0pt]

\item
\emph{epidemic} in that it uses a gossip
protocol~\cite{Birman:OSR:2007} to create a network-wide data
transmission and replication overlay. Instead of requiring a
management node to obtain data directly from a monitored node using an
end-to-end path, the data are transferred to and stored at
intermediate nodes from which the data are then also available. Our
algorithm for selecting peers from among neighbors at each cycle in
the gossip protocol is designed to use connectivity metrics and a
standard random selection process to account for the dynamic MANET
topology and limited connectivity.

\item
\emph{delay tolerant} in that it opportunistically uses whichever
neighbors happen to be in range at a gossip cycle, rather than relying
on a fixed topology to reach some desired (albeit unknown) end
point. This is an approach to moving data across an unstable network
that probabilistically guarantees that the data will be in reach of a
management node.

\item
\emph{intelligent} in that it is sensitive to the time-bounded and
time-sensitive nature of time-series data. Moreover, it minimizes the
data exchanged among peers through a synchronization and aggregation
algorithm that takes account of the history of past encounters and
exchanges.

\end{itemize}

\noindent
We have implemented our method in Java~EE and carried out an extensive
set of validation experiments using the US~Naval Research Laboratory's
Common Open Research Emulator (CORE)
Extendable Mobile Ad-hoc Network Emulator (EMANE)
facilities~\cite{5680218}. We compare the performance of our
harvesting method against
four data harvesting methods,
through a case study: the capture of \emph{service dependence data}~\cite{Novotny+:IEEE-TNSM:2015}.
The experiments explore a
range of network and system dynamics, with the primary dependent
variable being the accuracy of the dependence graphs produced from the
data as compared to the ground truth.

The novelty of the work presented in this paper is two fold: (i)~an
extension of ideas developed in the domain of epidemic protocols to
solve the problem of efficiently spreading \emph{time-series} data and
(ii)~doing so in the context of the \emph{unstable and unreliable
environment} of~MANETs. The method is specifically designed to address the needs of service-based systems, however, in generalized form it can be applied to harvest any type of time-series data in MANETs such as data collected from IoT sensors, system monitors and other sources.

We next review related work. In Section~\ref{sec:technique} we
present the distributed monitoring architecture and gossip protocol
underlying our method. Our experimental setup and evaluation are
detailed in Sections~\ref{sec:setup} and~\ref{sec:results},
respectively. We conclude with a summary of our observations and a
look at on-going and future work in Section~\ref{sec:conclusion}.

\section{Related Work}
  \label{sec:related}

The problem of maintaining the availability of data in a MANET
environment can be seen as a special case of the more general problem
of data sharing in that environment. From this point of view, there
are three broad areas of related work: peer-to-peer data access, data
replication, and epidemic protocols. While these areas are not
orthogonal, they represent the major foci of relevant work in the
MANET literature.

\para{Peer-to-peer data access.} Several techniques have been proposed
to create peer-to-peer overlays that can compensate for the dynamic
nature of a mobile
network~\cite{Heer+:PerComW:2006,Oliveira+:WoWMoM:2005}.  The most
advanced of these techniques are distributed hash tables (DHTs), which
are designed to provide fast access to distributed data in the face of
``churn'' in the set of peers.

A DHT primarily serves as an efficient look-up service (``index''),
not as a storage service \textit{per se}~\cite{Shah:MSN:2011}.
As such, DHTs help improve the availability of data only in terms of
finding where in the network the data, and possibly redundant copies
of those data, can be found, but not in directly addressing what can
happen if the nodes storing the data are themselves
unavailable. Moreover, although designed for churn, DHTs are not well
suited to environments that experience especially high rates of churn,
as can be the case in MANETs.

\para{Data replication.}  The goal of data replication is to increase
data availability by creating consistent replicas on multiple network
nodes~\cite{Derhab+:CST:2009, Padmanabhan+:VLDBJ:2008}.  Most
techniques make use of information provided by or known in advance
about the users of the
data~\cite{Bellavista+:ISCC:2005,Hara:2003,Hara:2004,916653}, aiming
to create replicas that are topologically near to consumers. We make
the weaker assumption that such information is unknown.  Similarly,
caching techniques~\cite{Atsan+:CN:2013,Fiore+:TVT:2011,Hara+:TMC:2006,Hao+:WIRLES:2005} 
replicate frequently used parts of data into temporary stores topologically
closer to users. However, since requests for data are infrequent, and
typically only refer to small subsets of the available data, caching
is ineffective.

Some replication techniques try to optimize placement so as to
minimize the amount of data movement, either by limiting the locations
where data can be created~\cite{Pacitti+:VLDB:1999} or by delaying and
bundling updates for transfer in
bulk~\cite{MoonCho:IADIS-AC:2004}. These techniques are essentially
transactional in nature, viewing the data as a database that is
subject to updates, whereas data in our framework is in the nature of
a time series.

Only a small number of data replication techniques explicitly address
the problem of disconnected
nodes. Huang~et~al.~\cite{Huang+:CIKM:2003} and Wang and
Li~\cite{WangLi:INFOCOM:2002} use mobility and data usage patterns to
place and redistribute replicas.
Hauspie~et~al.~\cite{Hauspie+:CNRS:2002} attempt to predict network
partitions, by measuring the quality of links and end-to-end paths, in
order to decide where best to place replicas. Although these
techniques do indeed increase data availability in partitioned
networks, they introduce substantial additional network overhead to
test links and discover paths.

\para{Epidemic protocols.}  An epidemic protocol (and the related
technique of gossiping) is a distributed algorithm that uses periodic
interactions between nodes to propagate data through a
network~\cite{Birman:OSR:2007,Dimakis+:IEEE:2010,Friedman+:SIGOPS:2007,Reina:AHN:2015}.
Particularly relevant to the problem addressed in this paper is that
the protocols do not require reliable communication links. The peers
(or neighbors) engaged in the protocol are usually chosen randomly.
In the context of MANETs, epidemic protocols are typically used in
conjunction with in-network computations, such as distributed signal
processing~\cite{RabbatNowak:IPSN:2004,Yu+:INFOCOM:2004}, to decrease
the overall network load by propagating only aggregate data values.
Our method makes heavy use of epidemic protocols and in-network
computations, where those computations are instead meant to
intelligently synchronize the time-series data traversing common
network paths rather than perform simple data aggregations.

\section{The Harvesting Method}
  \label{sec:technique}

We now introduce our harvesting method by first describing the basic
monitoring architecture and then detailing how the data are
disseminated through the network using a gossip protocol.

\subsection{Monitoring architecture}
  \label{sec:architecture}

At each (participating) node in the network, the harvesting method
makes use of two kinds of components and a local store. One component
is a \emph{monitor}, situated at a node, responsible for gathering
time-series data about the local state of the network and hosted
applications at that node. Monitors store the time-series data
according to \emph{time slots} of some length and use a sliding
expiration window such that only a limited history is maintained. A
time slot represents either a single measurement or an aggregate
series of measurements taken within the period of time represented by
the time slot. The time series thus consists of a continuous sequence
of time slots. What specific data are collected and how monitors
manage to gain access to those data are not of concern here, as this
is largely a domain-specific problem solvable in a variety of
well-studied ways; in the case study of Section~\ref{sec:casestudy} we
describe one particular mechanism and architecture.

Monitoring data are disseminated through the network using the second
kind of component, \emph{synchronization agents}, also situated at
each node of the network. As their name implies, the agents are
responsible for synchronizing the data shared with peer agents. In
particular, the agent at a node passes local monitoring data, together
with data received from other agents, to the node's
peers. Importantly, the agent also uses the local store to maintain a
backup (i.e., history) of data received from other agents. In this
way, the agents collectively form a network-wide data dissemination
overlay.

The ultimate consumer of monitoring data is an \emph{analysis element}
supporting a network and/or service management task of some sort. We
assume that the analysis element is itself located at a (mobile or
fixed) node in the network. In fact, there may be several such
analysis elements located across the network, but we do not assume
that the location of an analysis element is a globally known
property. Notice, however, that since the dissemination overlay formed
by the synchronization agents includes the node hosting the analysis
element, a query to a local store by the analysis element is all that
is required to retrieve network-wide monitoring data.

\subsection{Gossip protocol}
  \label{sec:gossip}

Synchronization agents employ a \emph{gossip protocol} to proactively
disseminate the monitoring data throughout the network. Unlike
previous gossip protocols, which are used mainly to compute and
propagate aggregate values~\cite{Birman:OSR:2007,Dimakis+:IEEE:2010},
our method is designed specifically to deal with time-series data
whose utility is both time bounded and time sensitive
(Section~\ref{sec:intro}).

\begin{figure}[t]
\centering

\includegraphics[width=0.6\columnwidth]{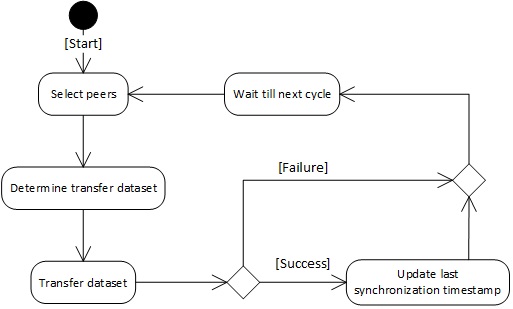}

\caption{The gossip cycle.}
  \label{fig:algorithm}
\end{figure}

At a high level, each agent repeatedly performs the following three
steps in cycles (Figure~\ref{fig:algorithm}):

\begin{enumerate}

\item \emph{Select peers.} The first step is to select the peers with
  which to synchronize. In order to maintain the simplicity, yet
  efficiency, of data synchronization under network mobility and
  wireless link unreliability, we use a random selection of the peers
  from candidates within a certain network distance.

\item \emph{Determine transfer datasets.} For each peer, a dataset to
  be sent is calculated based on changes in the data since the last
  successful synchronization with that peer. Two criteria are used to
  determine what dataset to send to each peer: data
  \emph{completeness} and data \emph{freshness}.

\item \emph{Transfer dataset.} The dataset for each peer is placed
  into a space-efficient data structure and sent to that peer. Since
  network links are unreliable, the agent records whether the transfer
  was successful by updating a synchronization timestamp associated
  with that peer.

\end{enumerate}

\noindent
We now describe each of these steps in detail.

\para{Select peers.}
The goal of peer selection is to ensure an even dispersion of the data
across the network in a simple yet effective manner. We use a
\emph{random}, \emph{memory-less} peer selection process,
 in which the peers are selected regardless of which and when
the data were previously sent~\cite{Jelasity+:ACM-TOCS:2007}, since it
is impractical to maintain a structured overlay topology across a
mobile network. Friedman~et~al.~\cite{Friedman+:SIGOPS:2007}
demonstrated that this strategy is effective for information
dissemination in~MANETs.

Using this process, each agent creates a candidate set of peers. The
candidate set is chosen based on a \emph{connectivity metric} such
that it contains only those with a high probability of successful data
transfer. It is well documented that as the hop distance between two
nodes increases in a MANET, the packet delivery ratio between those
nodes decreases
dramatically~\cite{DeCouto:2005:HPM:1150536.1150541}. Hence, we use
the hop distance metric, which can be conveniently obtained from the
local routing table, as the selection criterion. Specifically, a peer
is added to the candidate set if the hop distance is within a certain
threshold value (i.e., upper-limit parameter). A value of ``1'' is a \emph{neighbor} in the
network. Once the candidate set is established, a random subset of
nodes in the set is selected as the actual peers to be sent the
data. The number of selected peers is bounded by a configuration
parameter, as is the hop-distance upper-limit. 

\begin{figure}[t]
\centering

\begin{tabular}{@{}c@{\hspace{1em}}c@{}}
\includegraphics[width=0.3\columnwidth]{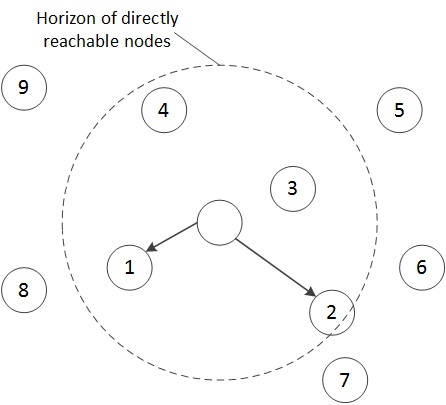}
&
\includegraphics[width=0.3\columnwidth]{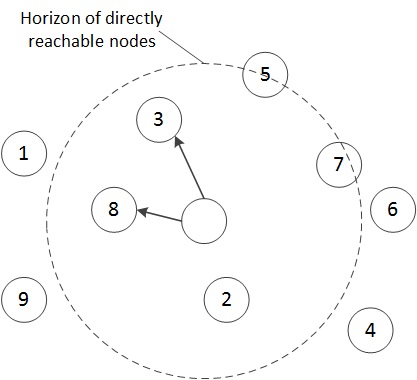}
\\
\textbf{(a)}
& \textbf{(b)}
\end{tabular}

\caption{Peer selections in two gossip cycles.}

 \label{fig:peer-selection}
\end{figure}

An illustrative example of the peer selection mechanism is shown in
Figure~\ref{fig:peer-selection}. In this example, the distance
threshold for nodes to be included in the candidate set is set to
1~hop. The upper bound of the number of peers is set to~2.

In the first cycle, shown in Figure~\ref{fig:peer-selection}a, the
candidate set will contain all nodes directly reachable from the
center node. The direct reachability is shown with the dashed circle.
The candidate set will thus contain nodes~\{1,2,3,4\}. From this
candidate set two peers \{1,2\} are randomly selected. In the next
gossip cycle, shown in Figure~\ref{fig:peer-selection}b, the positions
of the nodes have changed and now the candidate set contains nodes
\{2,3,7,5,8\} and the randomly selected peers are~\{3,8\}.

\begin{figure}[t]
\centering

\includegraphics[width=0.6\columnwidth]{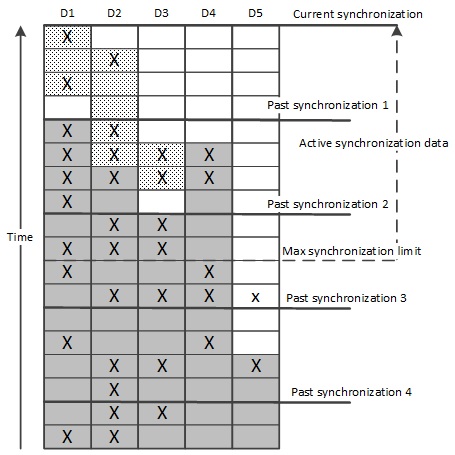}

\caption{Selection of time slots to include in a transfer dataset,
  sent from the agent to the peer. The transfer dataset (light shade)
  is extracted from the original data locally stored on the node of
  the source agent.}
  
  \label{fig:dataset-calculation}
\end{figure}

\para{Determine transfer dataset.}
The data stored locally by a synchronization agent are conceptually
organized into a table (Figure~\ref{fig:dataset-calculation}), with
columns for each time series. The rows contain values for each time
slot in the series. Once the peers are selected at the beginning of
the gossip cycle, each agent determines the dataset to
transfer to each peer individually. The goal of this process is to
propagate the data as efficiently as possible, as well as to keep the
data fresh. To do this, for each peer $i$, and for each time series
$d$, the source agent maintains (and updates) the most recent time
slot number, denoted by $t_s(d,i)$, in $d$ that was successfully
transferred to $i$. It then uses the following criteria to determine
which time slots of each time series to send to each peer:

\begin{enumerate}[leftmargin=*,label=(\alph*)]

\item Only the \emph{incremental changes} in each time series since
  the last successful synchronization to the peer are sent, i.e., only
  data in time slots $t > t_s(d,i)$ in time series~$d$ are sent to
  peer~$i$.

\item Time series data have an \emph{aging limit}, i.e., only the
  values in time slots no older than a maximum limit $T$ are
  transferred. This aging limit is a system configuration parameter,
  ensuring that obsolete data are not included in the synchronization
  process. Note, however, that eliminating data from a synchronization
  does not necessarily mean those data are dropped completely from the
  harvest, since agents could have received those data from other
  agents via other paths in the synchronization overlay.

\item To reduce communication overhead, certain empty time slots,
  which indicate the absence of monitoring data, are not transferred
  to peer~$i$. In particular, those that would appear at the beginning
  or end of the dataset are dropped, while those appearing between
  values are kept. We do this to simplify how a time series can be
  reconstructed at a peer. Of course, more sophisticated
  \emph{compression} techniques, both lossy and loss-less, could be
  considered, depending on how the monitoring data are to be used.

\end{enumerate}

\noindent
To enable the correct reconstruction of a time series, two additional
pieces of meta information are also transferred along with the values
of each time series: (i)~unique \emph{identifying information} for
each time series, defined and provided by the monitors that generate
the time series, and (ii)~the \emph{timestamp} of the last (newest)
time slot included in the dataset.

Figure~\ref{fig:dataset-calculation} provides an example of the
time-slot selection process at a given node as it attempts to
synchronize with a peer. The figure shows five time
series,~D1--D5. The symbol ``X'' represents the presence of some data
value in a slot, the dark shading indicates time slots that were
successfully sent to the peer in previous gossip cycles, and the
timings of the current and previous gossip cycles are shown as
horizontal lines.

The light shading indicates the time slots selected for transfer in
the current gossip cycle according to the criteria described
above. For time series D1, three time slots will be transferred; the
first and last are included because they contain some values, with an
empty time slot in between. For~D2, five time slots will be
transferred, but in this case some slots older than the last gossip
cycle are also included. These older slots were either received from
some other node after the last gossip cycle or were not successfully
transferred in a previous cycle.

Notice that the trailing (i.e., the most recent) empty slot is not
included in the transfer dataset. Similarly only two time slots are
included in the transfer dataset for D3, for which only those time
slots older than the last gossip cycle are not empty. Finally, no time
slot will be transferred for either D4 or D5: no new value is present
in D4 since the last successful synchronization, and the time slot of
new values in D5, while it was not previously transferred, is older
than the age limit.

\para{Transfer dataset.}
After the peer and transfer dataset have been determined, the dataset
is sent to that peer. However, since the network links can be
unreliable, the agent waits some predetermined timeout period for
confirmation of successful receipt.  Only once receipt is confirmed
does the agent advance the timestamp of the last successfully
synchronized time slot for that peer.

If, due to a network or other failure, the dataset transfer cannot be
completed or confirmed, the agent will abandon the synchronization
process with that peer in the current cycle. The confirmation timeout
period is a system parameter, obviously no longer than the time between
two successive gossip cycles, as this may lead to a race
condition in which the agent repeatedly attempts to send data that
might already have been received.

Note that the length of the time slots used by the gossip protocol may
not correspond to the length of the time slots used by the monitors.
Thus, the data maintained in the time slots by the monitors may be
aggregated and/or split in order to fit into the transfer time slots.
The aggregation of the data in the time slots used for transfer impacts
the overhead imposed on the network. This impact is explored in
Section~\ref{sec:results}.

Upon receipt of a dataset, an agent will reconstruct the transferred
portion of the time series (including any empty slots) and append it to
the time-series data in its local store. The data in the local store
are retained for a limited period of time, with obsolete data pruned
regularly to avoid extraneous use of resources.

\section{Evaluation Methodology}
  \label{sec:setup}

Our purpose in the evaluation is to understand the performance of the
harvesting method under various conditions. In particular, we are
interested in how well the method can improve the reachability (i.e.,
availability) of monitoring data in the face of network dynamics.

In this section we describe the method we use to evaluate our
approach. The method is based on a case study in which we experiment
with the problem of discovering service dependencies. After describing
the case study, we detail the tools, metrics, and scenarios used to
conduct the experiments.

\subsection{Case study: service dependence data}
  \label{sec:casestudy}

In previous work, we developed a technique for discovering dynamic
dependencies among the distributed components of MANET-hosted
applications that are structured as assemblies of
(micro)services~\cite{Novotny+:IEEE-TNSM:2015}. The technique suffered
from the problem that it assumes all nodes of interest are reachable,
on demand, from the node where the dependence analysis is to be
carried out. In fact, it is a common occurrence that not all nodes are
reachable, which significantly reduced the effectiveness of the
technique and inspired the design of our new harvesting method. In
this case study, we evaluate how well the new method can improve the
availability of the monitoring data and, thereby, the effectiveness of
the dependence discovery technique.

In service-based systems, a \emph{dependence} is a relation between
services defined by the message flow, called a \emph{conversation},
induced by a client request and normally ending with a response to
that request. (A dependence is also the relation between a client and
a service. Without loss of generality, we mainly focus here on
relations among services.)  When a dependence relation exists between
two services, one is considered the \emph{source} and the other the
\emph{target}. In general, sources issue requests (i.e., method calls)
on targets, thus defining a directionality to the dependence. Targets
are expected to provide replies (i.e., response messages) back to
sources.

\begin{figure}[t]
\centering

\includegraphics[width=0.6\columnwidth]{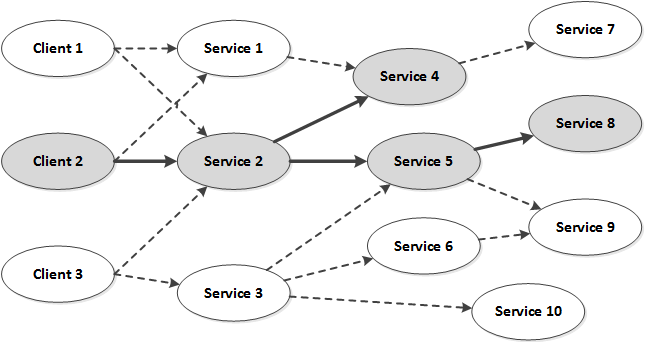}

\caption{Service dependence graphs.}
  \label{fig:service-assembly}

\end{figure}

A \emph{dependence graph} (DG) captures the run-time dependencies among
services and is the output of a dependence discovery analysis tool. A
DG can be used to represent the full set of dependence relations in the
system, or can be restricted to a subset of those relations.
Figure~\ref{fig:service-assembly} depicts a simple example of several
DGs rooted at clients. Highlighted in the figure are Client~2 and the
service instances it employs, both directly (Service~2) and
transitively (Services 4, 5, and~8).  A DG can be combined with network
and service failure data to perform global fault identification
tasks~\cite{Novotny+:SRDS:2012,Tati+:SPIE:2013}. For example, we can
probabilistically identify the root cause of a failed conversation.

To cope with network topology changes, service-based systems deployed
in MANETs make use of dynamic service binding
mechanisms~\cite{Mian+:IPC:2009}. This leads to time-varying
dependencies, which in turn are represented as a time series of data
points, each giving a snapshot of the dependencies at a given instant.
For meaningful use in dependence discovery, relevant and timely
dependence data must be available to the analysis tool.

\begin{figure}[t]
\centering

\includegraphics[width=0.6\columnwidth]{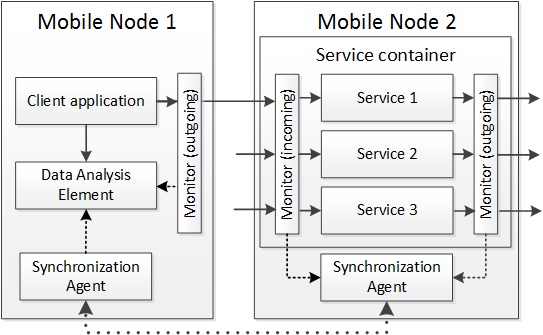}

\caption{Monitoring and harvesting dependence data. Note that only two
  of possibly many nodes are shown.}
  \label{fig:architecture}
\end{figure}

Consistent with the generic monitoring architecture
(Section~\ref{sec:architecture}), the dependence data are gathered by
local monitors. In this case, the monitors are deployed within service
containers to observe service-level message traffic
(Figure~\ref{fig:architecture}). The dependence data for a particular
client conversation, covering a specific time period, is provided for
the benefit of a dependence discovery analysis element that then
produces a corresponding~DG.  We assume that the dependence discovery
element is a component that can be hosted in any arbitrary node or
nodes (mobile or fixed) in the MANET, but that its specific location or
locations is not globally known.

Each time series represents the time-varying dependencies between a
source and target, in which entries for each time slot are Boolean
data about whether or not the given dependence occurred within that
time slot. When the monitor detects the occurrence of a dependence, it
signifies this by setting a 1-bit flag in the corresponding time
slot. It also records identifying information about the source and
target of the dependence. The set of time slots thus represents an
aggregated time series of dependencies. The set of relevant time slots
shifts as new time slots are added and obsolete ones removed,
reflecting the changes in dependencies.

\subsection{Experimental setup}
  \label{sec:expsetup}

As mentioned in Section~\ref{sec:intro}, we conduct our experiments in
the widely used
CORE\footnote{\url{http://cs.itd.nrl.navy.mil/work/core/}} network and
EMANE\footnote{\url{http://cs.itd.nrl.navy.mil/work/emane/}} mobility
emulator frameworks.  CORE provides a network experimentation
environment using the container-based virtualization facility of the
Linux platform, while EMANE provides real-time modeling of
wireless-link and physical-layer connectivity. This combination
provides high-fidelity, real-time emulation.  A detailed description
of the evaluation stack is provided elsewhere~\cite{
  NovotnyWolf:DTR-2016-7:2016}.

The application running within the MANET is built as a generic Web
service system based on Java~EE
Glassfish,\footnote{\url{https://glassfish.java.net/}} the reference
implementation of Java~EE's application platform, and Glassfish
Metro,\footnote{\url{https://metro.java.net/}} a reference
implementation of a standard Java Web services stack.  The system is
composed of two kinds of configurable components, a generic client
application and a generic Web service, structured as a 2-tiered
system. The first tier consists of client-facing, ``front-end''
services, while the second tier consists of interconnected ``back-end'' services. In our experiments, we use 50~clients, five front-end services, and 20
back-end services. When starting a conversation, each client invokes
a method selected uniformly at random from all methods provided by the
front-end services. The invocation triggers a cascade of message exchanges between the interconnected back-end services yielding a network traffic.

Monitors are implemented as \emph{Tubes} in the Metro framework.
Tubes are chained components, each responsible for part of the
processing of incoming and outgoing service messages.  Monitors
intercept incoming and outgoing messages of the clients and services
to extract the necessary information. The monitor extracts the
dependence fields from the intercepted messages and records
occurrences of dependencies.

Synchronization agents are implemented as Java applications that
repeatedly synchronize with other agents, receive data from remote
agents, and maintain in-memory backup stores of the received data. For
purposes of evaluation, synchronization agents also record information
about their activities in a trace file for later performance analysis
(e.g., measurements of network overheads, success rates of
synchronization attempts, and the like).

Finally, dependence discovery analysis elements, implemented as a
library, are used by clients to discover DGs for the conversations
they initiate (i.e., a series of message exchanges), by querying the
dependence information found in their local stores as populated by our
harvesting method.

\begin{table}
\centering

\begin{tabular}{@{}|l|r|@{}}
\hline
Number of nodes		& 50 \\
Mobility speed		& 3 - 6.6 km/h \\
WiFi standard		& 802.11b \\
WiFi unicast rate			& 11Mbps \\
WiFi multicast rate			& 1Mbps \\
Transmit power		& -15 dBm \\
Path loss mode		& 2ray \\
Routing protocol	& OLSR \\
Protocol stack		& TCP/IPv4 \\
\hline
\multicolumn{1}{c}{\mbox{}}\\
\end{tabular}

\caption{Network-layer parameters.}
  \label{tab:network-parameters}

\medskip

\begin{tabular}{@{}|l|r|@{}}
\hline
Number of clients		& 50 (one per node) \\
Number of services		& 25 \\
Size of dependence graph & 4 \\
Invokable methods per service	& 2 \\
Workload (client request rate)	& 30s \\
Number of service replicas & 5 \\
Response timeout		& 60s \\
\hline
\multicolumn{1}{c}{\mbox{}}\\
\end{tabular}

\caption{Service-layer parameters.}
  \label{tab:service-parameters}

\medskip

\begin{tabular}{@{}|l|r|@{}}
\hline
Monitor time slot length	& 0.1s \\
Transfer dataset time slot length	& 0.1s \\
Transfer dataset response time	& 60s \\
Maximum age limit of time slot		& 300s \\         
Maximum peer distance		& 1 hop \\
Number of gossip cycles & 0-32 \\
Number of peers & 1-10 \\
\hline
\multicolumn{1}{c}{\mbox{}}\\
\end{tabular}

\caption{Gossip protocol parameters.}
  \label{tab:agent-parameters}

\end{table}

Each (virtual) CORE node runs a client and/or a service, along with a
monitor, a synchronization agent, and an analysis element. (The
analysis element would not normally be deployed at all nodes, but we
do so here to give us maximum flexibility in evaluating the
dissemination of monitoring data.) The DGs are constructed on demand
by the discovery element. The graphs are rooted at a given client,
beginning at a given time instant, and for some time window. Each DG
is constructed for a particular conversation; thus, the time window
begins and ends with the start and end of that conversation.

In our experiments, one of the clients is chosen as the component
requiring the dependence data, such as when it needs to perform a
fault analysis as the initiator of a failed
conversation~\cite{Novotny+:SRDS:2012}. This reflects a realistic
use case in which the location of the analysis element cannot be
known \textit{a priori}.

To observe the effects of network dynamics, we use two different node
mobility patterns as scenarios. The first one, ``military'', consists
of a unit of 50 members, each carrying a mobile device. The members
are collected into several subunits, each of which moves as a whole
within an area of 2km~x~2km, exhibiting a group mobility driven by the
\emph{nomadic community mobility model}~\cite{Roy:2011}.  The second,
``firefighting'', represents a pattern of independent mobility of 50
members in a 1km~x~2km area, driven by the \emph{random waypoint
  mobility model}~\cite{Roy:2011}. Other network and service
parameters used in the experiments are summarized in
tables~\ref{tab:network-parameters}
and~\ref{tab:service-parameters}. Most notably, the mobile nodes move
at a walking-like speed in the range of 3 to 6.6km/h. We use WiFi
standard 802.11b and OLSR as a routing protocol.

The configuration of the gossip protocol for our experiments is
summarized in Table~\ref{tab:agent-parameters}. The lengths of the
monitor and transfer dataset time slots are both set to 0.1~seconds,
except as part of the last experiment presented in
Section~\ref{sec:results} in which we vary the transfer time slot from
0.1 to 10~seconds. The resolution of 0.1~seconds reflects the need for high precision of the captured dependencies necessary to construct accurate DG. 
The monitor waits at most 60 seconds for confirmation 
of successful receipt of the dataset. The maximum age of a time slot to be transferred is
limited to five minutes and the selection of candidate peers is
limited to network neighbors (i.e., 1-hop distant). The number of
gossip cycles within a given time period determines the frequency at
which each agent synchronizes the data in its local store with that of
its peers. The final parameter, number of peers, allows us to vary the
maximum number of other agents with which each agent shares the data
in its local store during each gossip cycle.







We collect our results from 40~minutes of execution after excluding
10~minutes of warm up. Each combination of parameters results in
thousands of conversations during the 40-minute execution. The results
given in the next section are averages over the data collected from
these conversations, where each conversation is then a statistical
sample subject to the random variables.

The primary evaluation question for our harvesting method is how well
the dependence data of each conversation~$C$ has been propagated
through the network after a certain period of time. More specifically,
we measure the quality of the harvest in terms of the ratio of
\emph{true positives} (TP) in the dependence analysis result, defined
as follows~\cite{Novotny+:IEEE-TNSM:2015,5691315}:

\begin{equation*}
{\mathit{TP\ ratio}} = \frac{| D(C) \bigcap GT(C)|}{|GT(C)|}
\end{equation*} 

\noindent where $D(C)$ is the set of discovered dependencies, $GT(C)$
is the set of ground-truth dependencies, and true positives are in the
intersection of these two sets (we assume $D(C)$ and $GT(C)$ are
non-empty). A good result for our method would be that it can transfer
as many dependencies of $C$ as possible, while not decreasing data
resolution due to aggregation.

The secondary evaluation question is the network overhead imposed by 
our harvesting method. We measure the overhead of a network node as an average of sum of all data sent and received by the synchronization agent hosted on the node within a unit of time (i.e. KB/s). The metric includes content (i.e. headers and payload) of control and data messages induced by the harvesting method in all peer-to-peer and client-to-peer exchanges. The dominant traffic in the network induced by the conversations between clients and services, as well as, the messages exchanged by the underlying routing protocol, are both excluded from the metric. 

\section{Experimental Evaluation}
  \label{sec:results}


We focus the evaluation of our delay tolerant harvesting method on
the following key issues:
\begin{enumerate}

\item impact of the synchronization frequency;
\item impact of the number of peers;
\item impact of constraining the number of peers;
\item comparison to four data harvesting methods;
\item tradeoff between overhead and precision.

\end{enumerate}

We study these issues in the context of the dependence discovery use
case.

\subsection{Impact of synchronization frequency}

The frequency at which peers attempt to disseminate monitoring data is
a fundamental tuning parameter in our method, as it induces a trade
off between communication overhead, on the one hand, and increased
availability of monitoring data on the other. To isolate the impact of
this parameter, we fix the delay between the end of the client
conversation and the start of the dependence analysis at five minutes,
and vary the number of gossip cycles within that five-minute period
from zero (no synchronization) to~32.
At each gossip cycle, we have each node select at most one peer node
at random from its neighbors in the network.

\begin{figure}[t]
\centering

\includegraphics[width=0.6\columnwidth]{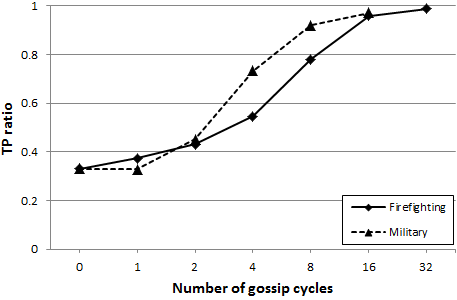}

\caption{Availability of monitoring data, as measured by TP ratio,
  with different numbers of gossip cycles.}
  \label{fig:ep-single-backup}

\end{figure}

We hypothesize that as the number of cycles increases in the period
between the end of the conversation and the start of the data
analysis, so too should the availability of the monitoring data,
resulting in an improved TP ratio. With no synchronization occurring
before the data harvesting, the only monitoring data available to the
harvesting node are those at its local monitor. However, with every
next cycle, the amount of data arriving from remote monitors should
increase.

The results are reported in
Figure~\ref{fig:ep-single-backup}. Initially, with no gossip cycles,
the only data available are from the client monitors and from the
locally hosted services. As the number of gossip cycles increases, the
nodes receive increasing amounts of monitoring data from remote
monitors via the gossip protocol.

Comparing the two mobility scenarios, the availability of the data
increases faster with an increasing number of gossip cycles under the
military scenario than under firefighting. This is because, in the
military scenario, the services involved in the conversations are with
high probability hosted on nodes within the same, relatively stable,
group of nodes as the client. On the other hand, in the firefighting
scenario, the services involved in conversations in general should be
drawn from greater distances, since the nodes are dispersed relatively
evenly over a large area, requiring the monitoring data to be
transferred between more nodes than in the military scenario and so
requiring more gossip cycles to achieve an equivalent TP ratio.
Eventually, with 32 cycles, the availability reaches 99.8\% in the
military scenario and 98.8\% in the firefighting scenario.

\subsection{Impact of number of peers}

We now investigate the impact of the number of peers selected in each
gossip cycle. Again, dependence analysis is started five minutes after
the conversation ends and the candidate peers are limited to
neighbors. We fix the synchronization frequency to one per minute,
i.e., four gossip cycles in that five-minute period. Our hypothesis is
that increasing the number of peers should increase the availability
of monitoring data.

In practice, the number of selected peers is bounded not only by the
upper-limit parameter, but also by the number of neighbors. Because
the size of the candidate neighbor set depends on the node density and
mobility of the network, the number of selected peers on average will
be somewhat lower than the configured limit. Furthermore, due to the
presence of poor-quality wireless links, the number of peers that
successfully receive the data can be smaller than the number of
selected peers. Especially when many peers are selected, congestion in
the wireless medium hampers message transmission.

\begin{figure}[t]
\centering

\begin{tabular}{@{}c@{}}
\includegraphics[width=0.6\columnwidth]{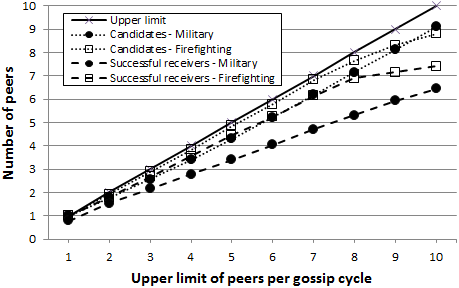}
\end{tabular}

\caption{Upper limit on peers, number of candidate neighbors, and
  number of peers successfully receiving data.}
\label{fig:ep-neighbors}

\end{figure}

This gap can be seen in Figure~\ref{fig:ep-neighbors}, which shows a
varying upper limit, the number of selected peers (``candidates''), and
the number of peers that
successfully receive the data under the two mobility scenarios. The
gap between these numbers widens as the upper limit increases. Notice
the subtle difference between the two scenarios. The gap consistently
increases in the firefighting scenario, where the nodes are evenly
dispersed. On the other hand, in the military scenario, there is
little gap between the upper limit and the number of selected peers
below a certain point (eight nodes), and then quickly widens beyond
that. This is due to the nature of the military scenario, where nodes
move in cohesive groups of certain sizes.

\begin{figure}
\centering

\includegraphics[width=0.6\columnwidth]{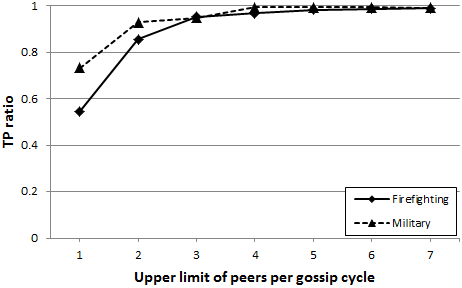}

\caption{Availability of monitoring data, as measured by TP ratio,
  with different upper limit on peers per gossip
  cycle.}
\label{fig:ep-multiple-backups}

\end{figure}

The availability of the monitoring data is reported in
Figure~\ref{fig:ep-multiple-backups}.  As expected, with increasing
numbers of peers per gossip cycle, the availability of the data also
increases. However, a threshold of about four peers is reached beyond
which there is little further gain. This is explained by the gap
between the upper-limit parameter and the number of peers successfully
receiving data, as discussed above. In principle, this threshold
should be taken into account in the configuration of the method to
balance overhead against the expected quality of the monitoring data
analysis result.

%
%
%

\subsection{Impact of constraining the number of peers}

In the next experiment we return to the synchronization frequency, but
remove the constraint on the number of peers.
Unlike in the previous experiments, the number of selected peers is
bounded only by the availability of the neighbors. Hence, the factor
that determines the inclusion of nodes is the node density and
mobility of the network. The dependence analysis is started five
minutes after the conversation ends and we vary the number of gossip
cycles within that period. We expect that increasing the number of
gossip cycles should increase the availability of monitoring data.

When the number of peers is unconstrained, data are propagated one hop
in each cycle from their source in all directions. Thus, the number of
cycles needed to make the data available to an analysis element
reflects the distance from the nodes on which those data are
collected. This distance, however, is impacted by the dynamics induced
by node mobility, as well as by the quality of the wireless
links. Furthermore, congestion in the wireless medium may have a more
significant impact on the transmissions than in the previous
experiments, since the data are sent in contiguous sequences to
multiple peers.

\begin{figure}[t]
\centering

\begin{tabular}{@{}c@{}}
\includegraphics[width=0.6\columnwidth]{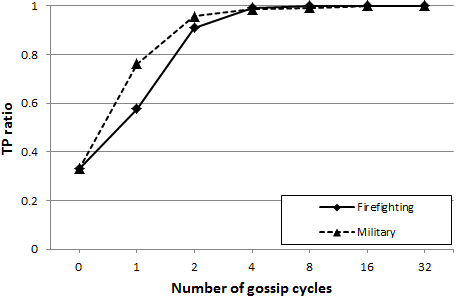}
\end{tabular}

\caption{Availability of monitoring data after four gossip cycles, as measured by TP ratio,
  with different numbers of gossip cycles and an unconstrained number
  of peers.}
  \label{fig:ep-unconstrained-backup}

\end{figure}

The results are reported in Figure~\ref{fig:ep-unconstrained-backup}.
In the firefighting scenario, where there is a relatively even
dispersion of nodes over a large area, the average number of
1-hop-distant peers is~7.2. In the military scenario, where nodes move
in relatively stable groups, the average number of 1-hop-distant peers
is~10.4. As with the one-peer-per-cycle configuration, increasing the
number of cycles increases the availability of the data. A threshold
of about two cycles is reached beyond which there is only a small gain
from further cycles. The higher number of 1-hop-distant peers in the
military scenario leads to a higher availability with same number of
cycles than in the firefighting scenario. However, to achieve a data
availability of 99\% or more, at least four cycles are needed in
either scenario to reach more distant nodes.

To compare the one-peer-per-cycle and unconstrained-peers
configurations, we evaluate the network overhead imposed by each
configuration to achieve a comparable data availability. In the
firefighting scenario, the one-peer-per-cycle configuration requires
32 cycles to achieve 98.8\% data availability, while the
unconstrained-peers configuration resulting in about 7.2 peers-per-cycle requires four cycles to achieve a
comparable 99.1\% availability. The overhead of the one-peer-per-cycle
configuration is 0.32~KB/s per node and 0.38~KB/s per node in the
unconstrained-peers configuration. In the military scenario, the
one-peer-per-cycle configuration requires 32 cycles to achieve a
99.8\% data availability, while the unconstrained-peers configuration
resulting in about 10.4 peers-per-cycle requires 8 cycles to achieve a comparable 99.3\% availability. The
overhead of the one-peer-per-cycle configuration is 0.19~KB/s per node
and about 0.29~KB/s per node in the unconstrained-peers
configuration. Hence, the unconstrained-peers configuration induces
overall about a 30\% to 50\% higher network overhead than the
one-peer-per-cycle configuration to achieve a comparable data
availability.

\subsection{Comparison to data harvesting methods}

To place our method in context, we next compare it to
two baseline methods: \emph{Gossip protocol} and \emph{DHT}, and to
two state of the art methods: \emph{DAFN} and \emph{SCALAR}.

\subsubsection{Gossip protocol}

A na\"{i}ve approach to harvesting would make use of a generic gossip
protocol that is not sensitive to the data already possessed by peer
nodes, and thus passes \emph{all} available data in \emph{each}
transmission regardless of past interactions.

We base the na\"{i}ve method on the design of a standard gossip
protocol~\cite{Jelasity+:ACM-TOCS:2007} and use a \emph{uniform
	random} selection of peers, \emph{push propagation} of data, with
\emph{no confirmation} of the success or failure of
transmissions. Nevertheless, rationally, we constrain the maximum age
of data to be passed. 

\subsubsection{DHT}
A similarly na\"{i}ve method to harvesting would make use of DHT with 
locations of time-series within the network, and attempt
to transfer data with direct, on-demand data harvesting approach. In that
approach, the analysis element attempts to communicate \emph{directly}
with individual monitoring nodes to obtain (``pull'') their data, rather
than using a gossip protocol to disseminate (``push'') the data. The
availability of the data is therefore limited by the network reachability of the
nodes.

\subsubsection{DAFN}

Dynamic Access Frequency and Neighborhood (DAFN) method~\cite{Hara+:TMC:2006} increases data availability with collaborative message relaying and placing data replicas on mobile
nodes based on data access frequency. Every node maintains a table of frequency
of accessing data. When node requests data, a request message is broadcasted to
its neighbors. Upon reception of the request message, node either responds with
data from its local cache or forwards the request to its neighbors. Upon
reception of the data, the node stores the data into the local cache. To
eliminate redundant data replicas between neighbor nodes, each connected set of
nodes elects a coordinator responsible for optimizing the allocation of
replicas. The coordinator monitors the allocation of replicas between neighbor nodes
and regularly prunes replicas on nodes with lower access frequencies.

\subsubsection{SCALAR}

Scalable data Lookup And Replication framework (SCALAR)~\cite{Atsan+:CN:2013}
is a scalable method of data lookup and passive replication in MANETs. The
method builds a dynamic virtual backbone between mobile nodes based on an
approximation of minimum connected dominating set within a set of connected
nodes. Thus, every node within a set of connected nodes is directly connected to
at least one node of the virtual backbone. The virtual backbone is used to
minimize the number nodes involved in relaying data request and response
messages within the network. A scalable data lookup protocol uses the virtual
backbone to relay data request and response messages between clients and data
sources. A reactive replication mechanism uses the data lookup protocol to
preload data from sources closer to clients. The mechanism preloads data based on
monitoring of request frequency and the distance between requester and source of
the data.

\subsubsection{Comparison of data availability}

\begin{figure}[t]
	\centering

	\begin{tabular}{@{}c@{}}
		\includegraphics[width=0.6\columnwidth]{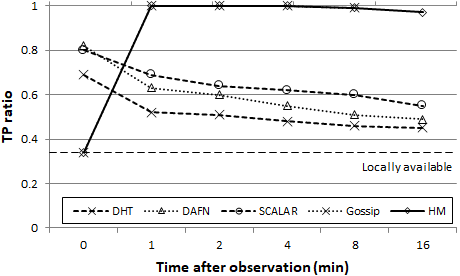}
		\\
		\textbf{(a)} \\[5pt]
		\includegraphics[width=0.6\columnwidth]{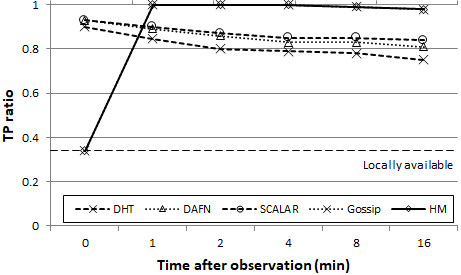}
		\\
		\textbf{(b)}
	\end{tabular}

	\caption{Comparison of TP ratio of dependencies with DHT, DAFN, SCALAR, Gossip protocol and our Harvesting method (HM) in the
		firefighting (a) and military (b) scenarios.}
	\label{fig:available-dependencies}
\end{figure}

Our aim in this experiment is to compare the data availability achieved by the methods under range of delays of starting data harvesting after the end of conversation.

Our method is configured with a single peer per cycle and a frequency of 32
cycles within the delay of data harvesting after the conversation and
the \emph{Maximum age limit of time slot} parameter is set to the length of the delay of data harvesting after the conversation.
Since the na\"{i}ve method uses the same fundamental gossiping mechanism, we
configure the method same as our harvesting method, with a single peer and
32 cycles within the delay period. 
In the DAFN method, we do not constrain the size of cache available for storing the time-series data replicas on nodes. Thus, the method’s decision on
reducing the number of data replicas between neighbor nodes is based solely on
the frequency of data access. Similarly, we do not constrain the size of the
cache available to data replicas in the SCALAR method. To achieve high data
availability, we configure the method to actively replicate incremental changes in
the time-series if there were more than one request
within the delay period.

Figure~\ref{fig:available-dependencies} shows the availability of
monitoring data using DHT, DAFN, SCALAR, Gossip protocol and our Harvesting method (HM), measured in terms of the TP ratio. The x-axis represents the delay between the end of a
conversation and the beginning of the harvesting.
Notice that the
general trend of data availability of DHT, DAFN and SCALAR methods follows the reachability of
nodes in the network shown in Figure~\ref{fig:reachability} of Section~\ref{sec:intro}. Initially, with no delay in issuing data harvesting requests after the conversation, data are highly available because the topology of the network has not significantly changed. Thus, methods which use the (``pull'') approach for obtaining data on request achieve high data availability. However, with increasing delay between end of conversation and the requests for data from clients, movements of nodes decrease the data availability.

The baseline DHT method, harvesting data directly on-demand over multiple hops achieves 90\% and 69\% TP ratios immediately after the conversation and declines to 75\% and 45\% TP ratios when requests for data are issued 16 minutes after the conversation in the military and firefighting scenarios, respectively. The DAFN method uses collaborative relaying of messages coupled with caching of obtained data on multiple replicas within the network. This approach increases the data availability to 93\% and 82\% TP ratios after the conversation and declines to 81\% and 49\% TP ratios at 16 minutes after the conversation. The SCALAR method based on relaying messages over virtual backbone, caching and limited (``push'') based replication yields similar data availability at 93\% and 80\% TP ratios after the conversation and declines to 84\% and 55\% TP ratios at 16 minutes after the conversation.

The baseline Gossip protocol and our Harvesting method both employ a (``push') approach to disseminate time-series data through the network. Both methods achieve very high data availability. Immediately after the conversation, before any data can be disseminated in the network, both methods achieve 34\% TP ratios in both scenarios, providing only data available in local stores. However, once the data start to be disseminated in the network, the availability increases to 99.9\% and 99.6\% TP ratios in the military and firefighting scenarios, respectively. The high availability somewhat declines to 98.5\% and 97.3\% TP ratios when requests are issued 16 minutes after conversation, caused by low frequency of disseminating data from migrating nodes.

\subsubsection{Comparison of network overhead}
\label{sec:comparison-network-overhead}
\begin{figure}[t]
	\centering

	\begin{tabular}{@{}c@{}}
		\includegraphics[width=0.6\columnwidth]{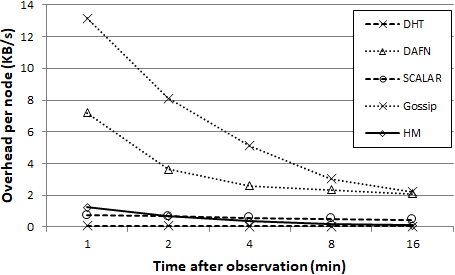}
		\\
		\textbf{(a)} \\[5pt]
		\includegraphics[width=0.6\columnwidth]{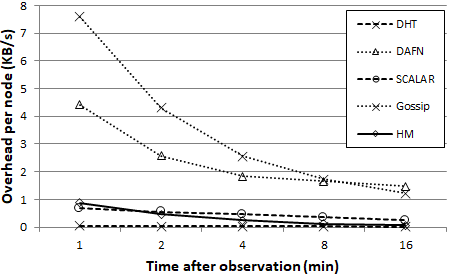}
		\\
		\textbf{(b)}
	\end{tabular}

	\caption{Comparison of transfer overheads of DHT, DAFN, SCALAR, Gossip protocol and our Harvesting method (HM) in the
		firefighting (a) and military (b) scenarios.}
	\label{fig:comparison-overhead}
\end{figure}

Here we compare the network overhead of the methods. We hypothesize that the DHT based method will induce only a small overhead, while the overhead of the Gossip protocol will be significantly higher due to its sending of the same data multiple times. The DAFN method, flooding the network with request and control messages will also induce significant overhead. The SCALAR method, utilizing caches and limiting the passing of messages to a subset of virtual backbone nodes will likely be more efficient. Our method should ideally induce significantly less overhead than the Gossip protocol and DAFN methods. However, since in our method the same data may be sent along different links at each cycle, the improvement will depend on the topology and connectivity properties of the network, and will certainly induce greater transfer overhead than the DHT. 
Note, that in this experiment, the delay between the end of conversation and beginning of harvesting varies among the scenarios, however, the total number of analyzed conversations is constant.

The results are reported in Figure~\ref{fig:comparison-overhead}. The DHT based method yields an overhead of only about 0.04~KB/s and 0.06~KB/s per node in the two scenarios with one minute delay and decreases to 0.02~KB/s and 0.04~KB/s per node with 16 minutes delay. The DAFN method yields significantly higher overhead, about 4.41~KB/s and 7.21~KB/s, and 1.47~KB/s 2.08~KB/s with one and 16 minutes delays respectively, and the SCALAR method yields moderate overhead of about 0.72~KB/s and 0.67~KB/s, and 0.34~KB/s 0.25~KB/s with one and 16 minutes delays respectively. 

The Gossip protocol yields substantially higher overhead of about 7.59~KB/s and 13.16~KB/s with one minute delay and decreases to about 1.22~KB/s and 2.21~KB/s with 16 minutes delay. Our method yields low about 0.87~KB/s and 1.24~KB/s and 0.07~KB/s and 0.11~KB/s transfer overhead to achieve the same levels of availability, placing it as expected between the DHT and DAFN as well as below SCALAR in all cases except in scenarios with one minute delay.

\subsubsection{Storage requirements}


The storage requirements are an artifact of the degree of data dissemination in
the network and the number of nodes holding data backups. The time-series data
are stored in the form of regularly pruned series of Boolean values. Overall,
the storage requirements of the harvesting methods are very small. Our method
uses all network nodes to proactively gossip all changes in the monitoring data
throughout the network with each node holding backup of the data received. In
the explored scenarios, the amount of data stored on the individual nodes was in
the order of tens of KB after data older than 20 minutes are pruned. Similarly,
DAFN uses all network nodes to relay data from monitors to clients, yet, with a
lower degree of data dissemination. The method further actively reduces the
number of data backups by eliminating neighbor replicas. The SCALAR method
limits the nodes participating in data transfer and holding data backups to
those of the virtual backbone. The DAFN and SCALAR methods required about 22 to
29\% and 8 to 12\% of storage space compared to our method, respectively. The
Gossip protocol achieves the same degree of data dissemination as our method and requires same storage space, while the DHT based method does not store any
backup data and therefore does not use any storage space.  

\subsubsection{Comparison summary}

The transfer of incremental changes of the time-series data is a challenging task for the data transfer methods. The DHT and the Gossiping protocol represent the two extremes of minimum overhead with low data availability and very high overhead with high availability respectively. The DAFN method with very high overhead provides only somewhat higher availability than DHT. The method is repeatedly flooding the network with requests for latest additions of the time-series data and makes low utilization of caches due to data becoming quickly obsolete. The SCALAR method provides somewhat higher availability than DAFN with lower overhead due to utilization of virtual backbone and caching reducing the number of messages requesting and disbursing latest additions of the time-series data. Our Harvesting method, designed specifically to continuously and efficiently transfer incremental changes in the time-series is yielding very high data availability with low network overhead.

\subsection{Tradeoff between overhead and precision}

As a final point of evaluation we look at the important issue of
overhead versus precision. 
The amount of data transmitted depends on the size
of the system (i.e., the number of time series), the configuration of
the method (i.e., gossip cycle frequency and number of peers per
cycle), and the length of the time slot within the time series. The
size of the system is an application-specific contextual property,
while the configuration settings reflect operational requirements. The
length of the data transfer time slot impacts overhead imposed on the
network, as well as the resolution (compression) of the transmitted
data. It thus impacts the precision of the resulting analysis applied
to the monitoring data.

To understand this effect for our case study, we use the ratio of
\emph{false positives} (FP)~\cite{Novotny+:IEEE-TNSM:2015,5691315} in
dependence graphs to indicate the impact of (im)precision in the
dependence data, defined as follows:
\begin{equation*}
{\mathit{FP\ ratio}} = \frac{| D(C) - GT(C)|}{|D(C)|}\nonumber
\end{equation*}
where as for the TP ratio defined in Section~\ref{sec:expsetup},
$D(C)$ is the set of discovered dependencies and $GT(C)$ is the set of
ground-truth dependencies, under the assumption that $D(C)$ and
$GT(C)$ are non-empty.

The FP ratio represents the fraction of dependence data not belonging
to a conversation and so erroneously included in the result; a high FP
ratio indicates poor precision. These irrelevant dependencies arise
from a combination of the monitors aggregating the dependence data
into time slots and the inherent behavioral effects of our harvesting
method. We hypothesize that increasing the length of the time slot
used in the transfer of data will decrease the communication overhead,
but increase the FP ratio.

\begin{figure}[t]
\centering

\includegraphics[width=0.6\columnwidth]{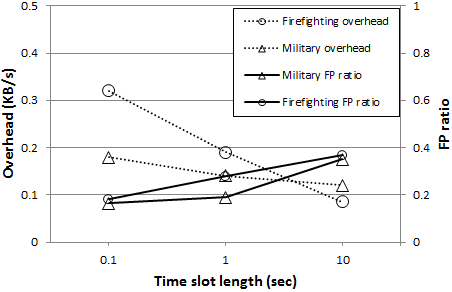}

\caption{Impact of the transfer dataset time slot length on FP ratio
  and overhead.}
\label{fig:resolution}

\end{figure}

We present our results in Figure~\ref{fig:resolution}, where we show
the impact of the time slot length on the overhead and FP ratio.
Consistent with the previous experiments, the method is configured
with a single peer per cycle and a frequency of 32~cycles within
period of five minutes, yielding a high availability.  The results
confirm our hypothesis: with a 0.1~second time slot length, the same
as that used by the monitor, the FP ratio is about 16\% and 18\% in
the military and firefighting scenarios, respectively, while the
overhead is 0.18~KB/s and 0.32~KB/s on average between pairs of
peers. When increasing the length of the time slot to 10~seconds, the
FP ratio increases to 35\% and 36\%, while the overhead decreases to
0.12~KB/s and 0.85~KB/s.

We observe that when the length of the time slot is decreased by 100
times, the FP ratio increases by only about two times, while the
overhead decreases by 33\% and 73\%. This suggests that for
applications able to tolerate a higher FP ratio, trading overhead over
higher FP ratio might be a viable option. Note that on average a node
sends only between 0.09 to 0.32~KB/s of data depending on the scenario
and the length of time slot.

\subsection{Summary of results}
  \label{sec:configuration}

The experimental results presented above establish the relationship
between the configuration parameters of our method (namely, the
synchronization frequency, the number of peers selected at each cycle,
and the length of the transfer dataset time slot), and its ability to
improve the availability of monitoring data.

In general, it is shown that selecting either a higher frequency with
a lower number of peers, or a lower frequency with a higher number of
peers, provide similar outcomes measured in terms of the availability
of the monitoring data at remote nodes. Furthermore, selecting a
shorter length for the transfer time slot provides higher precision
than a longer time slot. However, the data transfer overhead is
significantly higher with a shorter time slot length.

When considering network overhead, configurations using a lower number
of peers and a higher number of cycles achieves comparable data
availability with lower overhead, than configurations with a higher
number of peers and a lower number of cycles. 

In comparison to DHT, DAFN, SCALAR and Gossip protocol, our approach achieves
significantly higher data availability and lower overhead, especially in dynamic network
environments such as MANETs, where mobility can easily lead to nodes
becoming unreachable, yet their data remaining critical for analysis.

\section{Conclusion}
  \label{sec:conclusion}

We have presented a method to improve the availability of time-series
monitoring data for managing service-based systems.
To overcome the limited connectivity of the MANET nodes, the method
transfers the data over intermediate nodes in successive gossip
cycles. The method minimizes the network overhead caused by continuous
and repeated data transfer by storing local information about previous
transfers, imposing limits on the age of the data, and eliminating
irrelevant data. Through an extensive set of emulation-based
experiments, we have evaluated the capacity of the method to transfer
data from monitors to management nodes in two types of MANET
environments. Within the context of a case study, we demonstrated that
the method improves the results of management analysis tasks due to
the increased availability of monitoring data. Moreover, we have shown
how to tune the method to minimize the network overhead in the
resource-constrained MANET environment.

We have observed that the storage and communication overhead of the
method is relatively small. Of course, the actual amount of data
stored and transferred will depend on the nature of the time series of
interest. 
Putting this into context, the overhead is
negligible compared to the overhead of the service platform and
services hosted within it.


We have compared the method to four other data harvesting methods and shown that harvesting time-series data in MANETs is a challenging task in which the method performs significantly better in both, data availability and network overhead.


An important issue that we have yet to explore is how to select the
parameters when the fundamental nature of the network dynamics
changes. This is a challenging problem because a good configuration is
dependent upon a variety of factors, including: (i)~network topology
characteristics, such as node degree and network diameter, which
affect the data propagation speed critical to the epidemic process;
(ii)~node mobility characteristics, which affect the patterns of node
groupings and network partitioning; (iii)~urgency of the data, which
determines how quickly the data need to be available at remote nodes;
and (iv)~the communication bandwidth available for the epidemic
process.

Ideally, we would like the method to be self-configuring, such that it
can recognize the factors above and deduce appropriate configuration
parameters. There are several approaches to consider incorporating
into our design for this purpose:

\begin{itemize}

\item
a closed-loop process in which the current network status and the data
propagation rate are piggy-backed within the data synchronization
messages;

\item
an advanced peer-selection process that considers the history of past
dataset transfers and the full or partial knowledge of the current
dataset stored in the peer candidates, rather than a statistically
random selection process, as a way to increase the speed of data
dispersion within the network;

\item
a hybrid approach that combines the ``push'' epidemic data propagation
method with the ``pull'' on-demand method, where the pull would be
initiated only when the data from remote nodes are recognized to have
not yet been pushed all the way to the management node;

\item
a back-pressure-like protocol for data dissemination in which the
synchronization agents can adjust their synchronization rates
according to available communication bandwidth and the data
availability at other nodes; and

\item
a broadcasting based data dissemination approach to increase the rate of data dispersion in suitable cases (such as with high density of one hop-distant neighbors).

\end{itemize} 

\noindent
We are currently exploring these approaches in our on-going work,
which should result in an adaptive framework for data propagation and
harvesting in dynamic mobile networks.

\section*{Acknowledgment}

This research was sponsored by the U.S. Army Research Laboratory and
the U.K. Ministry of Defence and was accomplished under Agreement
Number W911NF-06-3-0001.  The views and conclusions contained in this
document are those of the author(s) and should not be interpreted as
representing the official policies, either expressed or implied, of
the U.S. Army Research Laboratory, the U.S. Government, the
U.K. Ministry of Defence or the U.K. Government. The U.S. and
U.K. Governments are authorized to reproduce and distribute reprints
for Government purposes notwithstanding any copyright notation hereon.





\section*{References}

\bibliographystyle{model1-num-names}
\bibliography{thispaper.bib}







\end{document}